\documentclass[aps,prc,preprint,superscriptaddress,showpacs,floatfix,nobibnotes]{revtex4}

\usepackage{bm}
\usepackage{graphicx}
\usepackage{longtable}

\begin{document}

\title{The thermal evolution of nuclear matter at zero temperature and definite baryon number density in chiral perturbation theory }

\author{Xiao-ya Li}
\affiliation{Department of Physics, Sichuan University, Chengdu,
610064, China}

\author{Xiao-fu L\"{u}}
\affiliation{Department of Physics,
Sichuan University, Chengdu, 610064, China} \affiliation{Institute
of Theoretical Physics, The Chinese Academy of Sciences, Beijing
100080, China} 

\author{Bin Wang}
\affiliation{Department of Physics, Sichuan University, Chengdu,
610064, China}

\author{Win-min Sun}
\affiliation{Department of Physics,
Nanjing University, Nanjing 210093, China} \affiliation{Joint Center
 for Particle, Nuclear Physics and Cosmology, Nanjing 210093, China}

\author{Hong-shi Zong}
\affiliation{Department of Physics,
Nanjing University, Nanjing 210093, China} \affiliation{Joint Center
 for Particle, Nuclear Physics and Cosmology, Nanjing 210093, China}

\date{\today}

\begin{abstract}
The thermal properties of cold dense nuclear matter are investigated with chiral perturbation theory.
The evolution curves for the baryon number density, baryon number susceptibility, pressure and the equation of state are obtained.
The chiral condensate is calculated and our result shows that
when the baryon chemical potential goes beyond $1150~\mathrm{MeV}$, the absolute value of the quark condensate decreases rapidly,
which indicates a tendency of chiral restoration.

\end{abstract}



\maketitle

\newpage

\parindent=20pt

\section{Introduction}
\begin{figure}[ht]
\centering\includegraphics[width=10cm]{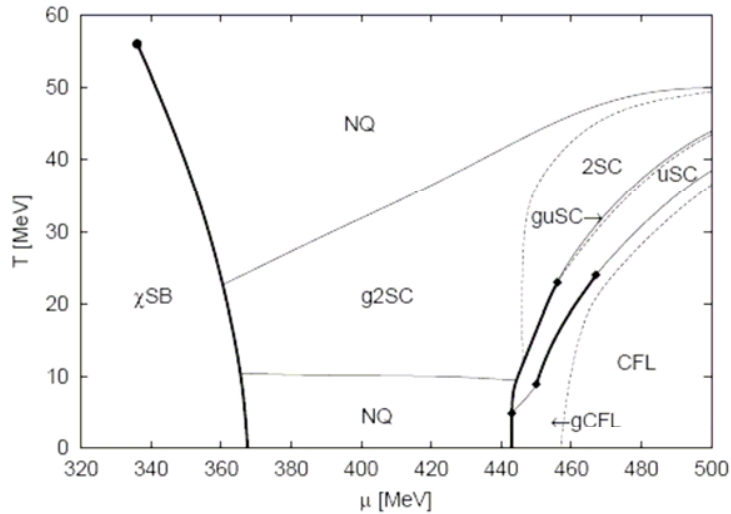}
\caption{\label{phasediagram} The phase diagram of QCD
for neutral strongly interacting matter given by Ref. \cite{ruster2006}.}
\end{figure}

The phase transition of strongly interacting matter is of great
interests to physicists.  The expected phase diagram of neutral strongly interacting matter given by Ref. \cite{ruster2006} in the framework of NJL model  is shown in Fig.~{\ref{phasediagram}}.
As the temperature and the chemical potential increase, there is a transition from hadron matter to quark gluon plasma (QGP). For small chemical potential, experiments at RHIC have found evidences for the existence of ideal-liquid-like QGP whose properties need further investigations \cite{RHIC05, RHIC;franz06, RHIC;dunlop07, phasediagram;munzinger08}, and LHC gives more information about QGP \cite{LHC;abreu07, LHC;enterric08}. Theoretically, lattice QCD  predicts that the transition of
hadron phase to QGP is a crossover with a critical temperature
in the range $150~\mathrm{MeV}\sim190~\mathrm{MeV}$ \cite{latticecrs;aoki2006,lattice;bornyakov05, latticeTcrs;cheng06,phasediagram;munzinger08}.
The mechanism of crossover has been studied
in many models \cite{crs;weise, crs;xum2008prl,crs:zhangzhao,crs;ejiri}.
For the low temperature and high chemical potential regime, our knowledge is quite limited. Experimentally, it is impossible to achieve such a condition in laboratories on earth. A natural laboratory holding such cold, highly compressed matter is
neutron star about which more observations and theoretical work
still need to be done \cite{Neutronstar;diener08,Neutronstar;weber05}. Theoretically, lattice QCD has difficulties in low temperature and high chemical potential regime. Model analyses predict that the phase transition is of first
order \cite{ruster2006, phasediagram;munzinger08, lowTNJL;buballa05,
lowT;Gerhold05}. As shown in Fig.~\ref{phasediagram}, for
strongly interacting matter with low chemical potential it is in
hadron phase and chiral symmetry is spontaneously broken. With
the increase of chemical potential, chiral symmetry is restored
and the hadron matter transits into QGP. When the chemical
potential increases to a certain point, QGP is expected to change
into a superconductivity phase \cite{Colorspc;alford07,
Colorspc;nardulli06}, the existence of which still needs to be verified.

To understand the thermal evolution of cold dense strongly-interacting matter,
one should first study how the baryon number density and the pressure change with the chemical potential and investigate the spontaneous breaking of chiral symmetry and its restoration where the key order parameter is the chiral condensate. However, at present an analysis of the thermal properties of cold dense strongly-interacting matter from first principles of QCD is not possible, so one has to resort to various nonperturbative QCD approaches and models. Chiral perturbation theory ($\chi$PT) \cite{Weinberg79, Gasser:85, Gasser:88,Leutwyler99} (for recent overviews, see Ref. \cite{ChPT;kubis07,ChPT;bernard07}) provides us an efficient method to investigate the problem in the regime where chiral symmetry is spontaneously broken.

In this paper, the thermal evolution of nuclear matter at zero
temperature is studied with $\chi$PT. In section II, the chemical
potential corresponding to the baryon number density of QCD is introduced and
is included into the effective Lagrangian of nucleon as an external
field. This method is analogous to the one used by
 Gasser and Leutwyler in Ref. \cite{Gasser:85, Gasser:88}. In section III, the
baryon number density, susceptibility and pressure are calculated
and the equation of state is obtained. In section IV, the thermal
evolution of quark condensate is considered. The evolution at
different temperature has been studied within $\chi$PT in Ref.
\cite{ChPTcondensate;martin06}. Our result at zero temperature is
consistent with the low temperature result obtained there, and shows the trend
of chiral restoration with the increase of chemical potential and
gives a restriction on the range of applicability of $\chi$PT results.
Section V gives the conclusions.

\section{Chiral Perturbation Theory at Zero Temperature and Definite Baryon Number density}
In QCD, the baryon number density corresponding to the conserved charge of $U(1)_B$ symmetry is $\frac{1}{3}q^+q$, where $q(x)$ denotes the quark field operator.
The baryon number of the system is given by $\langle \frac{1}{3}q^+q\rangle$ (the expectation value of the baryon number density operator on the vacuum state at definite $\mu$). At present, it is impossible to calculate $\langle \frac{1}{3}q^+q\rangle$ directly from the QCD Lagrangian. However, based on Weinberg's idea \cite{Weinberg79} that a hadron system could be described by an effective Lagrangian including all the possible terms consistent with the assumed symmetry principles, one can calculate the vacuum expectation value of quark operators effectively in $\chi$PT.

The calculation of $\langle\bar{q}\gamma_{\mu}q\rangle$,
$\langle\bar{q}\gamma_5\gamma_{\mu}q\rangle$,
$\langle\bar{q}q\rangle$ and $\langle\bar{q}\gamma_5q\rangle$ was
carried out by  Gasser and Leutwyler \cite{Gasser:85, Gasser:88} as
follows. Introduce four external fields $a_{\mu}$, $v_{\mu}$, s and p:
\begin{equation}\label{QCDL}
\mathcal{L}^{QCD}_{ext}=\mathcal{L}^{QCD}_{0}+\bar{q}\gamma^{\mu}(v_{\mu}+\gamma_5a_{\mu})q-\bar{q}\gamma^{\mu}(s-i\gamma_5p)q,
\end{equation}
where $\mathcal{L}^{QCD}_{0}$ is the QCD Lagrangian in the chiral
limit. The transformation properties of external fields are
determined through the invariance of $\mathcal{L}^{QCD}_{ext}$ under
local chiral transformation. In the end of the calculation, the scalar external field $s$ is set to be the mass matrix of quarks. According to Refs. \cite{Weinberg79, Gasser:85, Gasser:88},  if in the low energy regime the effective Lagrangian consists of all the possible terms, the results calculated from this Lagrangian are
equivalent to those calculated from QCD. So, the generating functional deduced
from this effective Lagrangian is equivalent to the one deduced from QCD. The
generating functionals are functionals of external fields which are included in the effective Lagrangian in a systematic way to ensure the local chiral symmetry.
Now the vacuum expectation value of quark operators $\bar{q}\gamma_{\mu}q$, $\bar{q}\gamma_5\gamma_{\mu}q$, $\bar{q}q$, $\bar{q}\gamma_5q$
can be obtained from the derivative of the generating functional of the hadron system with respect to the external fields corresponding to these operators.

To calculate the expectation value of the baryon number density $\langle \frac{1}{3}q^+q\rangle$ for cold dense nuclear matter,
we use the same method and introduce the chemical potential as follows.
Under local $U(1)_B$, the quark field transforms as
\begin{equation}
q\rightarrow e^{iB\theta(x)}q,
\end{equation}
where B is the baryon number operator. To ensure the local $U(1)_B$
symmetry of the system, we must introduce an external field $b_\mu$
which transforms as
\begin{equation}
b_\mu\rightarrow b_\mu+\partial_{\mu}\theta(x).
\end{equation}
Consequently the derivative acting on the quark field must be replaced by the canonical derivative
\begin{equation}
\partial_{\mu}q\rightarrow \partial_{\mu}q-B^qb_{\mu}q=\partial_{\mu}q-\frac{i}{3}b_{\mu}q.
\end{equation}
Set
\begin{equation}
b_\mu=\delta_{\mu0}\mu_b
\end{equation}
in the end of the calculation, then $\mu_b$ is the baryon chemical potential.
Here and in the following, in all calculations we assume flavor SU(2) symmetry, which means equal chemical potential and mass for u and d quarks.
The density of the conserved charge corresponding to $U(1)_B$ symmetry is the baryon number density operator $\frac{1}{3}\bar{q}\gamma_0q=\frac{1}{3}q^+q$.
Its vacuum expectation value can be calculated from the derivative of generating functional
\begin{equation}
\langle \frac{1}{3}q^+q\rangle=\frac{\partial Z[b_{\mu}]}{\partial b_0}|_{b_0=\mu_b,b_i=0}.
\end{equation}
In our approach, it is equivalent to replacing $b_0$ by $\mu_b$ and then performing the partial derivative with respect to $\mu_b$, and the result turns out to be
\begin{equation}
\frac{\partial Z[\mu_b]}{\partial \mu_b}.
\end{equation}
On the other hand, the nucleon field transforms as
\begin{equation}
\psi\rightarrow e^{iB\theta(x)}\psi
\end{equation}
under local $U(1)_B$ and the derivative of the nucleon field must be replaced by
\begin{equation}
\partial_{\mu}\psi\rightarrow \partial_{\mu}\psi-iB^Nb_{\mu}\psi=\partial_{\mu}\psi-i\delta_{0\mu}\mu_b\psi.
\end{equation}
Accordingly, the pion-nucleon effective Lagrangian $\mathcal{L}_{\pi N}$ for cold dense nuclear matter differs from the usual one in $\chi$PT \cite{ChPT;kubis07,ChPT;bernard07} in that the canonical derivative of the nucleon field has an additional term $-i\delta_{0\mu}\mu_b\psi$. The Lagrangian for the pure pion fields
$\mathcal{L}_{\pi}$ is the same as the one in the usual $\chi$PT, where the pions
are taken as Goldstone bosons corresponding to the spontaneous
breaking of chiral symmetry.

According to power counting, when calculating to the order of $O(p^4)$,
we need only to consider the first 4 orders of the effective
Lagrangian. For pure pionic Lagrangian, only the leading order term
$\mathcal{L}_{\pi}^{(2)}$ is needed:
\begin{equation}
\mathcal{L}_{\pi}^{(2)}=\frac{F_{\pi}^2}{4}Tr(\partial_{\mu}U\partial^{\mu}U^+)+\frac{F_{\pi}^2}{4}Tr(\chi
U^++U\chi^+),
\end{equation}
where $U$ denotes the Goldstone boson fields
\begin{equation}
U=exp(\frac{i\Pi\cdot\vec{\tau}}{F_{\pi}})
\end{equation}
with
\begin{equation}
\Pi=\left( \begin{array}{cc} \pi^0 & \sqrt{2}\pi_+ \\
\sqrt{2}\pi_- & -\pi^0 \end{array} \right)  \
\end{equation}
and $\vec{\tau}$ being the three Pauli matrices, $\chi=2B_0(s+ip)$ with $s$ and $p$ being the external fields introduced in Eq. (\ref{QCDL}).
In our calculation, $p=0$ and $s$ equals the mass matrix of u, d quarks. Under flavor $SU(2)_f$, $\chi$ can be expressed as
\begin{equation}
\chi=2B_0M_q=2B_0\hat{m}=M_{\pi}^2.
\end{equation}
For $\pi-N$ lagrangian, the terms contributing to our
calculation are
\begin{equation}
\begin{array}{rcl}\label{lagrangian}
\mathcal{L}_{\pi N}^{(1)}& =& \bar{\psi}(i\not{D}-m)\psi+\frac{1}{2}g_A\bar{\psi}\not{u}\gamma_5\psi,\\
\mathcal{L}_{\pi N}^{(2)}& =& c_1\langle \chi_+\rangle \bar{\psi}\psi-\frac{c_2}{4m^2}\langle u_{\mu}u_{\nu}\rangle (\bar{\psi}D^{\mu}D^{\nu}\psi+h.c.)
+\frac{c_3}{2}u_{\mu}u^{\mu}\bar{\psi}\psi+\cdots,\\
\mathcal{L}_{\pi N}^{(4)}& =&-\frac{e_1}{16}\langle\chi_+\rangle^2\bar{\psi}\psi+\cdots.
\end{array}
\end{equation}
where
\begin{equation}
\begin{array}{rcl}
D_{\mu}\psi& =& \partial_{\mu}\psi-i\delta_{0\mu}\mu_b\psi+\Gamma_{\mu}\psi,\\
\Gamma_{\mu}& =& \frac{1}{2}[u^+,\partial_{\mu}u],\\
u^2& =& U,u_{\mu}=iu^+\partial_{\mu}Uu^+.
\end{array}
\end{equation}
In Eq. (\ref{lagrangian}), $m$ denotes the nucleon mass in the chiral limit,  $c_1,c_2,c_3$ and $e_1$ are the low energy constants. $F_{\pi}$ is the pion decay constants. The values for the constants used in our calculation are shown in
Table~\ref{t1}. Here $c_4's$ and $e_2's$ are not shown because they are irrelevant to our calculation.

\begin{table}[ht]
\caption{\label{t1} The numerical values for $m$, $M_{\pi}$
$F_{\pi}$, $g_A$, $c_1,c_2,c_3$ and $e_1$ used in our calculation. $c_1,c_2$ and $c_3$  are the same as the ones used in Ref. \cite{ChPT;bernard07}. $m$ and $e_1$
are fitted through the nucleon mass.}. \vspace*{-6pt}
\begin{center}
\begin{tabular}{c|c|c|c|c|c|c|c}
\hline\hline
$M_{\pi}[MeV]$ & $m[MeV]$ & $F_{\pi}[MeV]$ & $g_A$ & $c_1[GeV^{-3}]$     & $c_2[GeV^{-3}]$    & $c_3[GeV^{-3}]$      & $e_1[GeV^{-2}]$ \\ \hline
 137           & 896    & 92.4           & 1.27    &$-0.90\times10^{-3}$ & $3.3\times10^{-3}$ & $-4.7\times10^{-3}$  & $1\times10^{-9}$ \\ \hline\hline
\end{tabular}
\end{center}
\end{table}

It can be seen that there are second and even higher order derivative terms of nucleon field in the effective Lagrangian,
such as the second term in $\mathcal{L}_{\pi N}^{(2)}$.
Due to the presence of those terms, the density of the conserved charge of $U(1)_B$ symmetry in nuclear system is not $\psi^+\psi$.
This is the reason why we introduce the baryon chemical potential at the quark level
instead of adding a $\mu_b\psi^+\psi$ term directly to the effective Lagrangian.
Besides, one may notice that the Feynman rules are different from those of the usual $\chi$PT
owing to the additional term in the canonical derivative of the nucleon fields.

\section{The thermal properties of cold dense nuclear matter}

In order to study the thermal properties of cold dense nuclear matter, we first calculate the baryon number density and its susceptibility.
According to the analysis in the above section, the baryon number density can be calculated in $\chi$PT from the effective Lagrangian of the nuclear system.
The pion-nucleon effective Lagrangian can always be written as
\begin{equation}
\mathcal{L}_{\pi N}=\bar{\psi}\hat{K}\psi,
\end{equation}
where according to Eq. (\ref{lagrangian}), $\hat{K}$ is of the form
\begin{equation}
\hat{K}=i\gamma_{\mu}\partial_{\mu}+\gamma_0\mu_b-m+\hat{O}
\end{equation}
with $\hat{O}$ denoting all the remaining terms of $\hat{K}$.
The baryon number density of the nucleon can be calculated to be
  \begin{equation}
  \begin{array}{rcl}
n(\mu_b)& = &\langle \frac{1}{3}q^+q\rangle = \frac{\partial}{\partial \mu_b}\int D\bar{\psi}D\psi DUe^{i\int d^4x\mathcal{L}_{eff}}\\
& = & \frac{\partial}{\partial \mu_b}\int DU(Det K)e^{i\int d^4x\mathcal{L}_{\pi}}\\
& = & \int DUTr(K^{-1}\frac{\partial}{\partial \mu_b}K)e^{i\int d^4x\mathcal{L}_{\pi}},
\end{array}
\end{equation}
where $\mathcal{L}_{eff}=\mathcal{L}_{\pi N}+\mathcal{L}_{\pi}$.
K and O denote the results of the Feynman path integral after integrating out the nucleon field. 
The most important contribution to $\langle \frac{1}{3}q^+q\rangle$ from K is
\begin{equation}
Tr(K^{-1}\gamma_0).
\end{equation}
The remaining term
\begin{equation}
Tr(K^{-1}\frac{\partial}{\partial \mu_b}O)
\end{equation}
comes from higher order terms in the effective Lagrangian which include the canonical derivative of the nucleon field,
such as the second term in $\mathcal{L}_{\pi N}^{(2)}$.
The contributions of such terms are far smaller than the leading term due to the smallness of $c_2$.
Therefore, we can neglect them in our later calculation and set
\begin{equation}
n(\mu_b)\doteq\int DUTr(K^{-1}\gamma_0)e^{i\int d^4x\mathcal{L}_{\pi}}.
\end{equation}
After performing the Feynman path integral on the pion fields U, $K^{-1}$ becomes the propagator S of the nucleon field.
Then the baryon number density is
\begin{equation}
n(\mu_b)=\frac{-i}{(2\pi^4)}\int dp^4Tr(\gamma_0S)
\end{equation}

It should be noted that the nucleon propagator is different from the usual one
because in the Feynman diagram the momentum of nucleon is always added with a term $\delta _{\mu0}\mu_b$. If the calculation is restricted to order $O(p^4)$, i.e. we only consider the diagrams shown in Fig. 2, then the self-energy is the same as what was shown in Ref. \cite{Leutwyler99} except that the momentum $p$ of nucleon field is replaced by $p'=\{p_0+\mu_b,\vec{p}\}$, wherever it appears.
The explicit expression for the self-energy is
\begin{equation}
\begin{array}{rcl}
\Sigma & =&-4c_1M^2+\Sigma_a+\Sigma_b+\Sigma_c+e_1M^4+O(p^5),\\
\Sigma_a& =&\frac{3g_A^2}{4F_{\pi}^2}(m+\not p')\{M_{\pi}^2I+(m-\not p')\not p'I^{(1)}\},\\
\Sigma_b& =&\frac{3M_{\pi}^2\Delta_{\pi}}{F_{\pi}^2}\{2c_1-\frac{p'^2}{m^2d}c_2-c_3\},\\
\Sigma_c& =&-4c_1M_{\pi}^2\frac{\partial\Sigma_a}{\partial m},
\end{array}
\end{equation}
where d=4, $\not p'=\not p+\gamma_0\mu_b$,
\begin{equation}
\begin{array}{rcl}
I & =& -\frac{1}{8\pi^2}\frac{\alpha\sqrt{1-\Omega^2}}{1+2\alpha\Omega+\alpha^2}ArcCos(-\frac{\Omega+\alpha}{\sqrt{1+2\alpha\Omega+\alpha^2}})
-\frac{1}{16\pi^2}\frac{\alpha(\Omega+\alpha)}{1+2\alpha\Omega+\alpha^2}(2ln\alpha-1),\\
I^{(1)}& =&\frac{1}{2p'^2}\{(p'^2-m^2+M_{\pi}^2)I+\Delta_{\pi}\},\\
\Delta_{\pi}& =&\frac{M_{\pi}^2}{8\pi^2}ln\frac{M_{\pi}}{m}
\end{array}
\end{equation}
and
\begin{equation}
\begin{array}{rcl}
\alpha& =&\frac{M_{\pi}}{m},\\
\Omega& =&\frac{p'^2-m^2-M_{\pi}^2}{2mM_{\pi}}.
\end{array}
\end{equation}
In fact, $\Sigma_{a,b,c}$ coincide with the corresponding one in
Ref. \cite{Leutwyler99}, except that the nucleon momentum $p$ is replaced
by $p'$. The renormalization scheme used here is the one
introduced by Leutwyler in Ref. \cite{Leutwyler99} based on the
"infrared regularization", which we denote as IR.
The baryon number density can now be calculated to be
\begin{equation}
n(\mu_b)=\frac{-i}{(2\pi^4)}\int d^4 p Tr(\frac{\gamma_0}{\not{p'}-m-\Sigma(\not{p}')}),
\end{equation}
\begin{figure}[!htb]
\centering
\includegraphics[scale=1,angle=0]{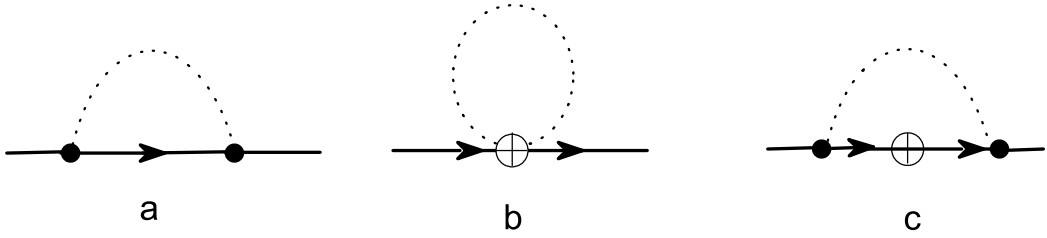}
\caption{\label{selfenergy} The one-loop diagrams contributing to the self-energy of the nucleon.
The crossed vertices denote contribution from $\mathcal{L}^{(2)}_{\pi N}$}
\end{figure}
Remember that the chemical potential is introduced in Euclidean space,
 $p_0$ is imaginary and the integral is of the form
\begin{equation}
\int d^4 p=\int d^3\vec{p}\int^{i\infty}_{-i\infty}dp_0.
\end{equation}
Besides, note that the baryon number density vanishes at zero chemical potential, so
if we let $p'_0=p_0+\mu_b$, then
\begin{equation}\label{integral}
n(\mu_b)=\frac{-i}{(2\pi^4)}\int d^3\vec{p}(\int^{i\infty+\mu_b}_{-i\infty+\mu_b}dp'_0-\int^{i\infty}_{-i\infty}dp'_0)Tr(\frac{\gamma_0}{\not{p'}-m-\Sigma(p')}).
\end{equation}
The integration path in the complex plane would be closed as was shown in Fig.~{\ref{integralpath}}.
\begin{figure}[!htb]
\centering
\includegraphics[scale=1,angle=0]{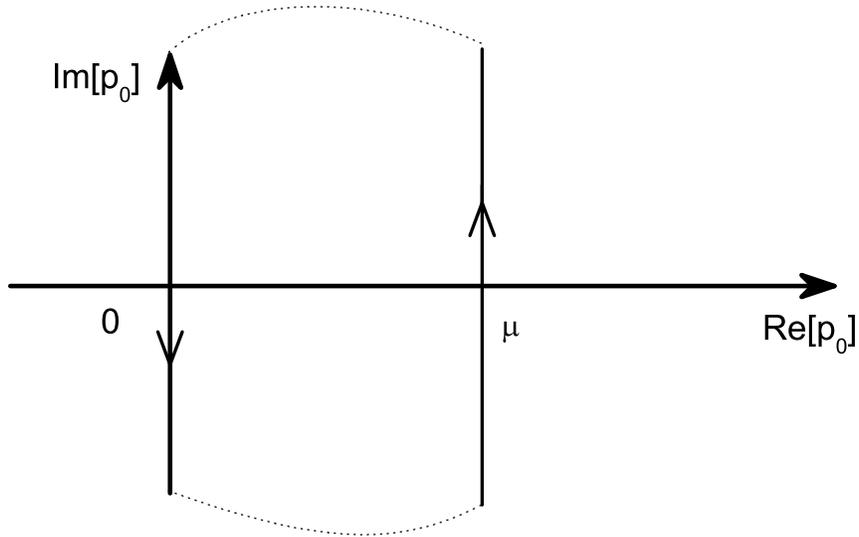}
\caption{\label{integralpath}
The integration path of $p_0$ for the baryon number density of the nucleon in Eq. (\ref{integral})}
\end{figure}
According to Cauchy integral formula, the value of the integral is determined by
the analytical properties of the integrand in the region encircled
by the integration path. However, a direct application of the IR
results \cite{Leutwyler99} in the full complex $p_0$ plane is
illegal. In IR, the loop integral H is divided into two parts, the
singular part I and the regular part R, and the R part is dropped
out in renormalization because it does not fit the infrared
singularity of the integral. Considering that the analysis of the
singularity is based on an expansion of the nucleon momentum around
the nucleon mass $m_N$, when the nucleon momentum goes far beyond $m_N$,
the expansion is illegal and so the regular part R can not be dropped
out. In the regime where the momentum of the nucleon is much
larger or smaller than $m_N$, the result of the loop integral should
be $H=I+R$, not I.

\begin{figure}[!htb]
\centering
\includegraphics[scale=1,angle=0]{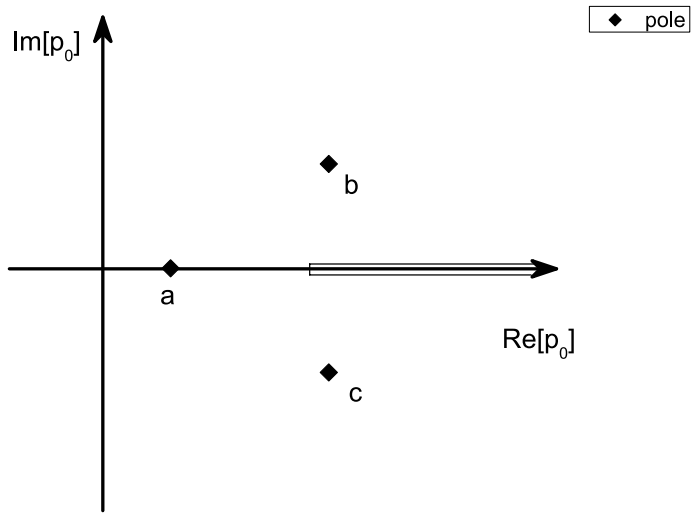}
\caption{\label{pole}
The poles and the branch cut of the integrand in Eq. (\ref{integral}) in the complex $p_0$ plane.
The positions of the three poles are at $p_0=\sqrt{\vec{p}^2+p_n^2}$, where n denotes a, b and c.
$p_n$ for each pole is $p_a=938$, $p_b=1152+i 337$, $p_c=1152-i 337$.
The branch cut starts at d where $p_0=\sqrt{\vec{p}^2+(m+M_{\pi})^2}$.}
\end{figure}

According to above consideration,
the integrand in Eq. (\ref{integral}) has three poles and one branch cut in the complex $p_0$ plane with the real part of $p_0$ positive and not larger than 1200~MeV.
The regime beyond 1200~MeV is out of the consideration of $\chi$PT because there the spontaneous breaking of chiral symmetry is expected to be restored.
Then the poles and the branch cut relevant to our calculation are shown in  Fig.~{\ref{pole}}.
Obviously, the pole denoted as "a" on the real axis corresponds to the ground state of the nucleon (N938),
and the conjugated pairs "b" and "c" can be explained as the excited modes of the nucleon.
The brach cut starts at $p_0=\sqrt{\vec{p}^2+(m+M_{\pi})^2}$.

\begin{figure}[!htb]
\centering
\includegraphics[scale=0.8,angle=0]{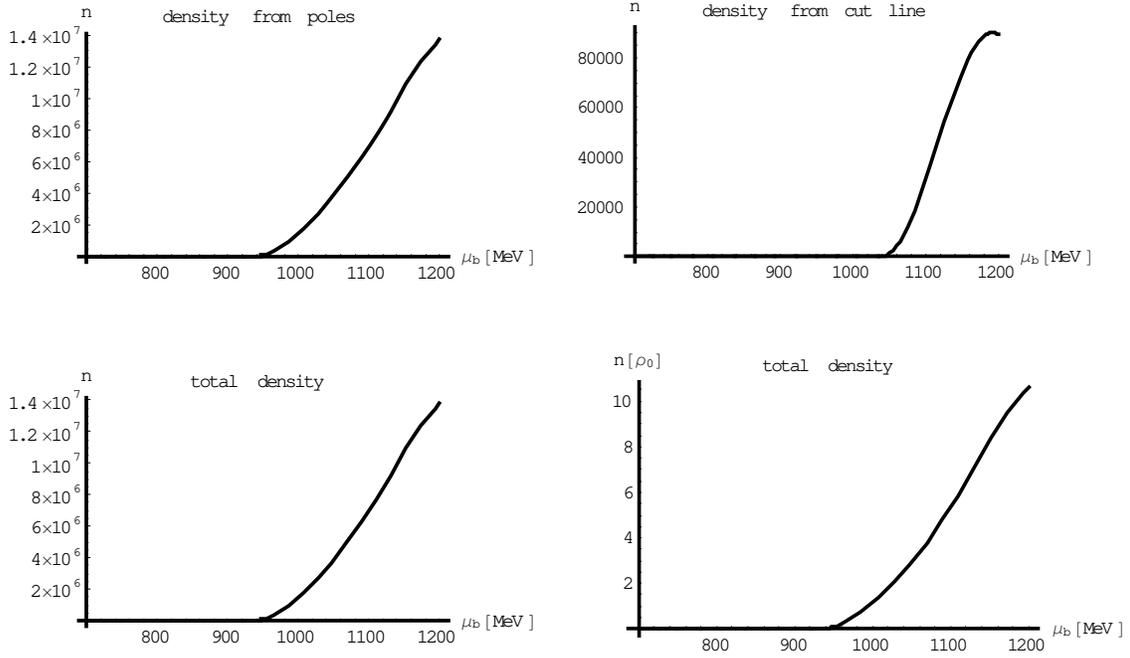}
\caption{\label{density}
The curves for the baryon number density of cold dense nuclear matter. The left top shows the contributions from poles.
The right top shows the contribution from the branch cut.
The left bottom shows the total density and the right bottom shows the total density in the unit of the saturated nuclear density $\rho_0$.}
\end{figure}

\begin{figure}[!htb]
\centering
\includegraphics[scale=0.8,angle=0]{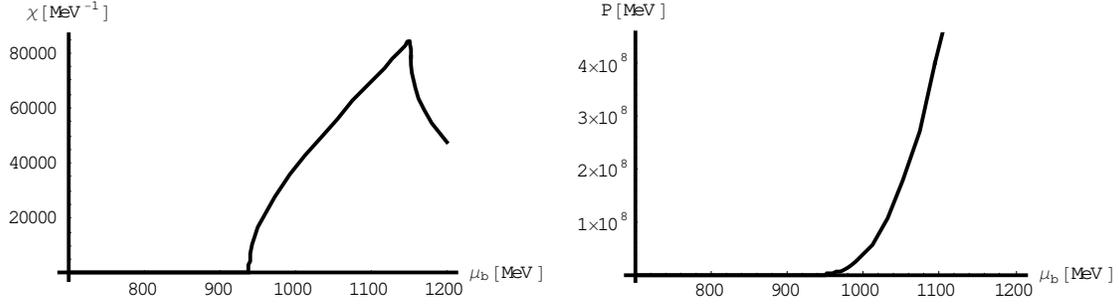}
\caption{\label{chiandp}
The curves for the pressure and the baryon number susceptibility of the cold dense nuclear matter.
The left shows the baryon number susceptibility. The right shows the pressure.}
\end{figure}

\begin{figure}[!htb]
\centering
\includegraphics[scale=0.8,angle=0]{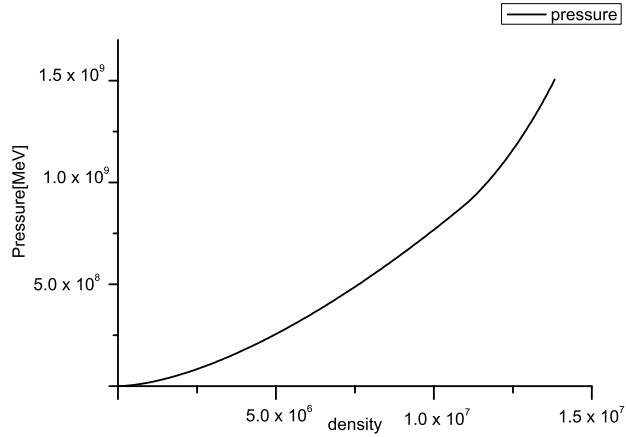}
\caption{\label{stateeq}
The curve for the equation of state of cold dense nuclear matter.}
\end{figure}

The results for the baryon number density are shown in Fig.~{\ref{density}},
where the contributions from poles and branch cut and the total contribution are shown respectively.
The susceptibility of baryon number density $\chi$ is calculated to be
\begin{equation}
\chi(\mu_b)=\frac{\partial}{\partial\mu_b}n(\mu_b).
\end{equation}
The pressure P of the nuclear matter can be calculated through the relation
\begin{equation}
n(\mu_b)=\frac{\partial P}{\partial\mu}
\end{equation}
which comes into existence when the temperature of the system stays the same.
The results of $\chi$ and P are shown in Fig.~{\ref{chiandp}}.
The $P-n$ curve for the equation of state is shown in Fig.~{\ref{stateeq}}.

The results indicate that the baryon number density $n(\mu)$ and pressure P of the nucleon are zero for baryon chemical potential $\mu_b$ smaller than the mass of the nucleon 938~MeV. That is, $\mu=938~\mathrm{MeV}$ is a singularity. This result agrees qualitatively with the general conclusion of Ref. \cite{Halasz:phaseofQCD}. In that reference, based on a universal argument, it is pointed out that the existence of some singularity at the point $\mu=\mu_0$ and $T=0$ is a robust and model-independent prediction.  A recent model calculation using the rainbow approximation of the Dyson-Schwinger approach \cite{Zong:EOS} also support this. When  $\mu_b$ is larger than 938~MeV, $n(\mu)$ and P increase with the increase of baryon chemical potential.
And when $\mu_b$ is larger than 1152~MeV, the rate for the increase of $n(\mu)$ and P is slowed down. The peak in the curve of susceptibility indicates a second order phase transition around $n(\mu)=1150~\mathrm{MeV}$, which is due to the exited modes of nucleon.
Physically, the critical point for $n(\mu)$ and P to go from zero to a nonzero value should be smaller than 938~MeV due to contribution from the binding energy.
However, our calculation cannot reveal such contribution because the pion loops are not considered. It can be expected that when calculating to higher order, the results would be better and the quenching effect would be revealed.

Remember that $\chi$PT is available only when chiral symmetry is spontaneously broken.
As we have indicated in the introduction,
with the increase of chemical potential there is a transition from hadron phase to QGP for which the chiral symmetry is restored.
So, the applicability of above results should be limited in a certain range.
In order to analyze the restoration of chiral symmetry and determine the range of applicability of our results,
we need to calculate the quark condensate $\langle \bar{q}q\rangle$ which characterizes the spontaneous breaking of chiral symmetry. This will be done in the next section.

\section{The evolution of the quark condensate}
According to $\mathcal{L}_{ext}^{QCD}$ in Eq. (\ref{QCDL}),
the quark condensate $\langle\bar{q}q\rangle$ can be calculated from the derivative of the generating functional with respect to the corresponding external fields
\begin{equation}
\langle\bar{q}q\rangle=\frac{\partial}{\partial s}\int D\bar{\psi}D\psi DUe^{i\int d^4x\mathcal{L}_{eff}}|_{s=M_q}.
\end{equation}
As the baryon chemical potential is not included in $\mathcal{L}_{\pi}$, the pure pion loop does not contribute to our calculation.
By analogy with the deduction in the last section, the result comes out to be
\begin{equation}\label{condensateitg}
\frac{-i}{(2\pi^4)}\int d^4 pTr(S\frac{\partial\Sigma}{\partial\hat{m}}),
\end{equation}
where $\hat{m}$ denotes the mass of u and d quarks under flavor SU(2) symmetry.
The integral is performed in the same way as in Eq. (\ref{integral}).
\begin{equation}\label{}
\int d^4 p=\int d^3\vec{p}(\int^{i\infty+\mu_b}_{-i\infty+\mu_b}dp'_0-\int^{i\infty}_{-i\infty}dp'_0)
\end{equation}
with the integration path shown in Fig.~{\ref{integralpath}.
However, since the quark condensate does not vanish at zero chemical potential,
the result of the integral in Eq. (\ref{condensateitg}) is not $\langle\bar{q}q\rangle$ but instead $\langle\bar{q}q\rangle-\langle\bar{q}q\rangle_0$,
where $\langle\bar{q}q\rangle_0$ denotes the quark condensate at zero chemical potential.

The curve for  $-(\langle\bar{q}q\rangle-\langle\bar{q}q\rangle_0)$ is shown in Fig.~{\ref{condensatem}}
\begin{figure}[!htb]
\centering
\includegraphics[scale=0.8,angle=0]{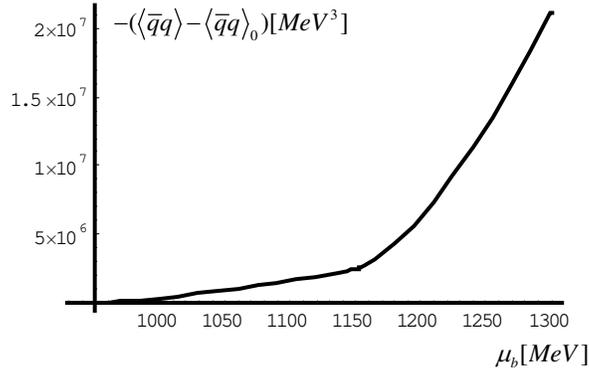}
\caption{\label{condensatem}
The curve for the evolution of $-(\langle\bar{q}q\rangle-\langle\bar{q}q\rangle_0)$.}
\end{figure}
and the curve for the quark condensate $-\langle\bar{q}q\rangle$ is shown in Fig.~{\ref{condensate}},
\begin{figure}[!htb]
\centering
\includegraphics[scale=0.8,angle=0]{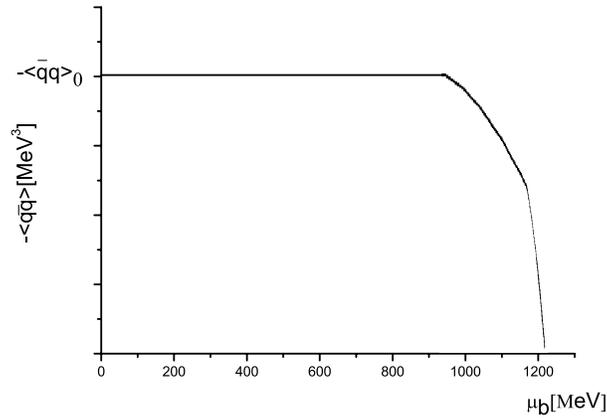}
\caption{\label{condensate}
The curve for the evolution of $-\langle\bar{q}q\rangle$.}
\end{figure}
 $\langle\bar{q}q\rangle$ remains the same as the quark condensate at zero baryon chemical potential, i.e., $\langle\bar{q}q\rangle=\langle\bar{q}q\rangle_0$. A recent model calculation using the rainbow approximation of the Dyson-Schwinger approach  gives the same result \cite{Zong:ConDS}.  For $\mu_b$ larger than $m_N$, the absolute value of the quark condensate tends to decrease as $\mu_b$ increases,
and the decrease speeds up when $\mu_b$ is larger than 1150~MeV,
which means that the spontaneous breaking of chiral symmetry tends
to get restored. Therefore, the $\chi$PT result is applicable only
for $\mu_b$ smaller than 1150~MeV where the spontaneous
breaking of chiral symmetry still works. Our results are consistent
with the HB$\chi$PT result at low temperature in
\cite{ChPTcondensate;martin06}, where the evolution of the quark
condensate is investigated for different temperatures. And in principle our
results are also consistent with that in of Refs. \cite{quarkcondensateshao06,
Leichang}. Although $\chi$PT is not applicable, the
point at which the hadron-QGP transition occurs can be estimated to be around
1150~MeV, which is consistent with the result of
$\mu_q=\frac{1}{3}\mu_b\sim365~\mathrm{MeV}$ expected in NJL model \cite{ruster2006}.

\section{Conclusion}
The thermal properties of cold dense nuclear matter are analyzed and
the results for the evolution curve of the baryon number density, the baryon number
susceptibility and the pressure are shown in Fig.~{\ref{density},
\ref{chiandp}}. The curve for the equation of state is shown in
Fig.~{\ref{stateeq}. An analysis of the quark condensate shows that chiral symmetry tends to get restored for chemical potential larger than
1150~MeV. Our results are applicable only in the
range $\mu_b<1150~\mathrm{MeV}$, where the spontaneous breaking of chiral
symmetry approximately holds. Our results show that the transition
of nuclear matter to QGP occurs at around 1150~MeV, which is
consistent with the prediction in \cite{ruster2006}. To understand
the full thermal properties of cold dense strongly interacting
matter, we need to investigate the region above the phase
transition point using a method, such as the DSE method in
\cite{DSE;zong05A, DSE;zong05B, DSE;zong05C}, which is applicable for cold
dense quark matter. As is indicated in section III, the
contribution of binding energy and the quench effect are not
included in our calculation. To improve our result we need to make
the calculation to order of $O(p^6)$.

\begin{acknowledgements}
We have benefited greatly from discussions with L. Chang, Y. X. Liu,
C. D. Roberts. This work is supported in part by the Key Research Plan of Theoretical
Physics and Cross Science of China (under Grant No. 90503011), the National Natural Science Foundation of China (under Grant No. 10575050) and the Research Fund for the Doctoral Program of Higher Education (Grant No. 20060284020).
\end{acknowledgements}

\end{document}